# Numerical investigation of effective nonlinear coefficient model for coupled third harmonic generation

Zihua Zheng[1], Ziwen Tang[1,2], Zhiyi Wei[3,4], Jinghua Sun[1,5]

[1]*Institute of Ultrafast Optics and Photonics, Dongguan University of Technology. Dongguan 523808, China*
[2]*Guangdong Provincial Key Laboratory of Nanophotonic Functional Materials and Devices, South China Normal University, Guangzhou 510006, China*
[3]*Beijing National Laboratory for Condensed Matter Physics, Institute of Physics, Chinese Academy of Sciences, Beijing 100190, China*
[4]*Songshan Lake Materials Laboratory, Dongguan 523808, China*
[5]*sunjh@dgut.edu.cn*

**Abstract:** In this paper, the optimal solution of effective nonlinear coefficient of quasi-phase-matching (QPM) crystals for coupled third harmonic generation (CTHG) was numerically investigated. The effective nonlinear coefficient of CTHG was converted to an Ising model for optimizing domain distributions of aperiodically poled lithium niobate (APPLN) crystals with lengths as 0.5 mm and 1 mm, and fundamental wavelengths ranging from 1000 nm to 6000 nm. A method for reconstructing crystal domain poling weight curve of coupled nonlinear processes was also proposed, which demonstrated the optimal conversion ratio between two coupled nonlinear processes at each place along the crystal. In addition, by applying the semidefinite programming, the upper bound on the effective nonlinear coefficients $d_{eff}$ for different fundamental wavelengths were calculated. The research can be extended to any coupled dual $\chi^{(2)}$ process and will help us to understand better the dynamics of coupled nonlinear interactions based on QPM crystals.

## 1. Introduction

As one of the most important inventions of our time, laser technology has very important applications in numerous fields, such as precision measurement [1, 2], laser processing [3], fundamental physics [4, 5], and quantum information [6, 7]. However, the wavelengths of lasers depend on the energy level structure of gain mediums, which occupy limited ranges with respect to the whole optical spectrum. Nonlinear optical frequency conversion techniques are essential to extend the applications [8]. Quasi-phase matching (QPM) [9-11] is a widely used conversion technique, in which, the sign of the second-order susceptibility is conventionally changed periodically in a piece of nonlinear crystal (lithium niobate, for example) to obtain a periodically poled structure, offering an additional reciprocal lattice vector to satisfy the momentum conservation condition. Generally, a periodically poled lithium niobate (PPLN) crystal can only satisfy a single $\chi^{(2)}$ process in a narrow bandwidth. However, in practical applications, multi-process frequency conversion in a single QPM crystal is expected for rich spectrum availability with high efficiency, low cost and decent stability, which cannot be realized by conventional PPLN crystals.

To solve this problem, successive researchers have broken through the limitation of periodically poling design and proposed quasi periodically poled lithium niobate (QPPLN) [12], chirped periodically poled lithium niobate (CPPLN) [13, 14], and aperiodically poled lithium niobate (APPLN) [15,16] crystals. Through flexible modulation of the domain lengths, multiple nonlinear processes and wide spectrum bandwidth can be realized in a single QPM crystal. For multi-process frequency conversion, APPLN can achieve higher conversion efficiency than conventional cascaded PPLNs [26]. Optimizing the poling direction and the length of the each domain in APPLN can be obtained by maximizing the effective nonlinear coefficient $d_{eff}$ through methods such as simulated annealing (SA), genetic algorithm [17] and even quantum annealing [18]. Taking the CTHG [9, 15] that contains both second harmonic generation (SHG, $\omega+\omega=2\omega$) as well as sum frequency generation (SFG, $\omega+2\omega=3\omega$) as an example, for different fundamental wavelengths $\lambda_1$, what are the distribution modes of the poling directions of the corresponding optimal CTHG crystal? How does the effective nonlinear coefficient vary with the fundamental wavelengths? Does $d_{eff}$ have a theoretical upper bound? These have been less addressed in previous studies. The understanding of these questions will help us to understand the conversion mechanism of CTHG better.

In this paper, we convert the effective nonlinear coefficient model to the Ising model [19], and use SA algorithm to search the optimal domain distributions of APPLN and calculate the corresponding CTHG $d_{eff}$ with crystal lengths as 0.5mm and 1mm , respectively, when fundamental wavelength varys from 1000nm to 6000nm. We also propose a method to reconstruct the poling weight curves of domain distributions from the obtained APPLNs. In addition, by applying the semidefinite programming [21], we calculated the upper bound of the $d_{eff}$ for different fundamental wavelengths.

## 2. Methodology for CTHG crystal design

The QPM technique introduces reciprocal lattice vectors by changing the sign of the crystal's nonlinear coefficients within a coherence length, which compensates the phase mismatch during nonlinear frequency conversion process. As shown in Fig. 1, the APPLN crystal is divided into equal-sized domains, each can have either "up" or "down" poling direction. (It should be noted that in some articles, the term "domain" have different definition which refers to the length of continuous poling in the same direction, in that case each domain may be unequal-sized.) Optimized arrangement of the directions of the domains can achieve the most efficient nonlinear frequency conversion.

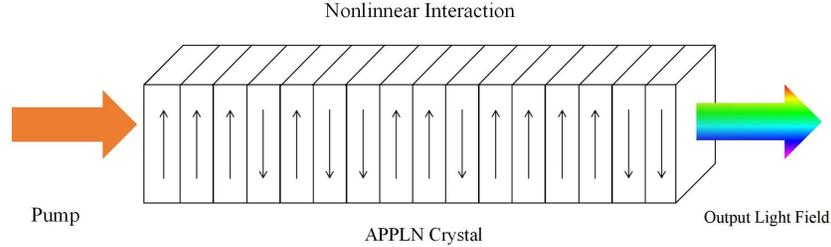

Fig. 1. Schematic diagram of the APPLN structure. A lithium niobate crystal is divided into multiple equal-sized domains, and each domain takes either "up" or "down" poling direction. The consecutive domains with the same direction form a real poling domain for QPM crystal manufacturing process.

The CTHG process contains both SHG and SFG nonlinear processes. Under plane-wave approximation, the coupled wave equations [22] are

$$\frac{dE_1}{dz} = \frac{i\omega_1 d_{33}}{n_1 c} d(z) E_1^* E_2 \cdot \exp(i\Delta k_1 z) + \frac{i\omega_1 d_{33}}{n_1 c} d(z) E_2^* E_3 \cdot \exp(i\Delta k_2 z)$$

$$\frac{dE_2}{dz} = \frac{i\omega_2 d_{33}}{2n_2 c} d(z) E_1^2 \cdot \exp(-i\Delta k_1 z) + \frac{i\omega_2 d_{33}}{n_2 c} d(z) E_1^* E_3 \cdot \exp(i\Delta k_2 z) \qquad (1)$$

$$\frac{dE_3}{dz} = \frac{i\omega_3 d_{33}}{n_3 c} d(z) \cdot E_1 E_2 \cdot \exp(-i\Delta k_2 z)$$

where $i$ is imaginary unit, $E_1$, $E_2$ and $E_3$ are complex amplitudes of fundamental, second and third harmony, respectively. $\Delta k_1 = n_2\omega_2/c - 2n_1\omega_1/c$ and $\Delta k_2 = n_3\omega_3/c - n_1\omega_1/c - n_2\omega_2/c$ are the phase mismatches of the SHG and SFG processes, respectively. $d_{33}$ is the nonlinear coefficient of the lithium niobate crystal in the 33 direction, $c$ is the speed of light in a vacuum, and $\omega_1$, $\omega_2$, and $\omega_3$ are the frequencies of the fundamental wave, second harmonic wave, and third harmonic wave, respectively. $n_1$, $n_2$, and $n_3$ are the corresponding refractive indices, respectively. All simulations performed at 80°C in this paper. $d(z)$ is the poling direction of the domain at position $z$, where $d(z) = +1$ means "up" and $-1$ means "down". Under the small signal approximation, the CTHG conversion efficiency is given by [15]

$$\eta_{THG} \equiv \frac{I_3}{I_1} = \frac{144\pi^4 |d_{33}|^2 I_1^2 L^4}{c^2 \varepsilon_0^2 \lambda^4 n_1^3 n_2^2 n_3} (d_{eff})^2, \qquad (2)$$

where $I_1$ is the fundamental wave intensity, $I_3$ the third harmonic wave intensity, $L$ the crystal length, $\lambda$ the fundamental wavelength, $\varepsilon_0$ the vacuum permittivity, and $d_{eff}$ the effective nonlinear coefficient defined as [15]

$$d_{eff} = \left| \frac{2}{L^2} \int_0^L dz\, e^{i\Delta k_2 z} d(z) \times \int_0^z dx\, e^{i\Delta k_1 x} d(x) \right|. \qquad (3)$$

For given pump and signal wavelengths, $\Delta k_1$ and $\Delta k_2$ can be obtained from Sellmeier's equation [23]. Equation (2) shows that to maximize $\eta_{THG}$, we need to find an optimal set of domain arrangements $d(z)$ that maximizes the $d_{eff}$. The crystal is divided into $N$ domain with equal length $\Delta x = L/N$. Discretize the equation (3) to obtain

$$\begin{aligned} d_{reff}^{THG} &\approx \frac{2}{L^2} \left| \sum_{m=1}^{N} \int_{(m-1)\Delta x}^{m\Delta x} dz\, e^{i\Delta k_2 z} d(z) \times \sum_{n=1}^{m-1} \int_{(n-1)\Delta x}^{n\Delta x} dx\, e^{i\Delta k_1 x} d(x) \right| \\ &= \frac{2}{L^2} \left| \sum_{m=1}^{N} \frac{1}{i\Delta k_2} [e^{i\Delta k_2 m\Delta x} - e^{i\Delta k_2 (m-1)\Delta x}] d(m) \times \sum_{n=1}^{m-1} \frac{1}{i\Delta k_1} [e^{i\Delta k_1 n\Delta x} - e^{i\Delta k_1 (n-1)\Delta x}] d(n) \right|. \end{aligned} \qquad (4)$$

Equation (4) can be abbreviated as

$$d_{eff} = \left| \sum_{m=1}^{N} \sum_{n=1}^{m-1} J(m,n) d_m d_n \right|. \qquad (5)$$

Here, $d_m \equiv d(m) \in \{+1, -1\}$ is the poling direction of the number $m$ crystal domain, and $J(m,n)$ is the coupling coefficient which is expressed as

$$J(m,n) = \frac{2}{L^2 \Delta k_1 \Delta k_2} (e^{i\Delta k_2 m\Delta x} - e^{i\Delta k_2 (m-1)\Delta x}) \cdot (e^{i\Delta k_1 n\Delta x} - e^{i\Delta k_1 (n-1)\Delta x}). \qquad (6)$$

Simulations show that for different $d_m$, complex value of $\sum_{m=1}^{N}\sum_{n=1}^{m-1} J(m,n) d_m d_n$ starts at zero in the complex plane, when the maximum mode lengths in different directions are approximately equal. To simplify the computation, we define the objective function $f$ as the projection of $d_{eff}$ on the real number axis,

$$f(d_1, d_2, \ldots, d_N) = \sum_{m=0}^{N} \sum_{n=1}^{m-1} J_0(m,n) d_m d_n, \qquad (7)$$

where the real coupling coefficient

$$J_0(m,n) = \frac{2}{L^2 \Delta k_1 \Delta k_2} \{\cos(\Delta k_2 m\Delta x) - \cos[\Delta k_2 (m-1)\Delta x]\} \{\cos(\Delta k_1 n\Delta x) - \cos[\Delta k_1 (n-1)\Delta x]\} \qquad (8)$$

The Ising model [19, 24] plays an important role in statistical physics [27], which defined as $H = -\sum_{ij} J_{ij} \sigma_i \sigma_j$, where spins $\sigma_i, \sigma_j \in \{+1, -1\}$ and coupling coefficient $J_{ij} \in \mathbb{R}$. From equation (7), it can be seen that the objective function $f$ is an Ising model [19, 24], whose value is determined by the poling direction of each crystal domain, and the maximum value can be obtained by using SA algorithm.

For an APPLN with $N$ calculation crystal domains, there are $2^N$ different combinations, and there is no effective algorithm yet to find the global optimal solution in reasonable time using classical computers. Article [18] broke through the limitation of classical computers and proposed a method of quantum annealing algorithm to search for the global optimal solution, which is a very promising algorithm in the field of APPLN design. However, the current quantum computers have limited number of quantum bits, which hampers the accuracy of crystal domain design. For these reasons, this paper still chooses SA algorithm to search for the optimal solution. In this way, the solution might be a local optimum, but it is the most feasible and effective solution under the current computer development. Unless otherwise specified, "optimal" in this paper refers to the best result obtained by our search in a reasonable time.

For the fundamental wavelength $\lambda_1$, its second harmonic wavelength and third harmonic wavelength are $\lambda_2 = \dfrac{\lambda_1}{2}$ and $\lambda_3 = \dfrac{\lambda_1}{3}$, respectively. The phase mismatch $\Delta k_1 = k_2 - 2k_1$ and $\Delta k_2 = k_3 - k_1 - k_2$ are for SHG and SFG, respectively, where $k_i = \dfrac{2\pi}{\lambda_i} \cdot n_i$, in which $n_i$ is refractive indices ($i$=1,2,3).

## 3. Optimal $d_{eff}$ at different wavelengths

In our simulation, the fundamental wavelength ranges from 1000nm to 6000nm with a 10nm interval for crystal length $L$ as 0.5mm and 1mm, respectively. $L$ to be limited in 1mm is benefit to computing time and reasonable for the applications in femtosecond region. The SA simulated maximum value of $d_{eff}$ is shown in Fig. 2.

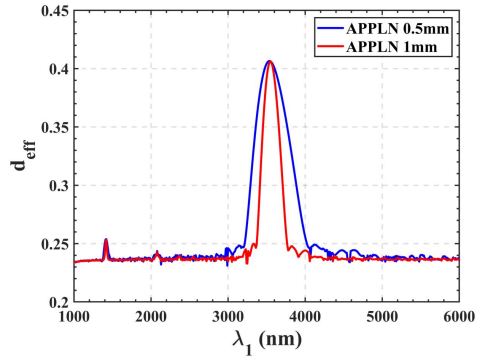

Fig. 2. The values of the $d_{eff}$ corresponding to the optimal APPLN crystal for fundamental wavelengths of 1000nm-6000 nm when the crystal length L are 0.5 mm (blue) and 1 mm (red), respectively.

In Fig.2, the effective nonlinear coefficient $d_{eff}$ at most of fundamental wavelengths are approximately equal to a fixed value of about 0.235, and the curves show peaks at the wavelengths near 1420nm, 2080nm, and 3550nm. The largest peak appears near $\lambda_1 = 3550$nm. In order to understand the mechanism, we calculated the phase mismatches $\Delta k_1$ and $\Delta k_2$ for SHG and SFG at different fundamental wavelengths, and the results show that the peaks of the curves of $d_{eff}$ in Fig.2 are locating at the positions where $\dfrac{\Delta k_2}{\Delta k_1}$ are integers as show in Fig.3. Especially, at $\lambda_1$=3550nm where the $d_{eff}$ curves have the highest peaks, $\Delta k_1$ equals to

$\Delta k_2$, which referred as *degenerate point*. Fig.2 and Fig.3 show that for high efficient CTHG, it's better to choose the fundamental wavelengths in the cases where $\frac{\Delta k_2}{\Delta k_1}$ are integers. Of course, the locations of these special wavelengths depend on different crystal materials. It can be of benefit to obtain supercontinuum generation if the fundamental wavelengths are chosen at the degenerate point of QPM crystals [25].

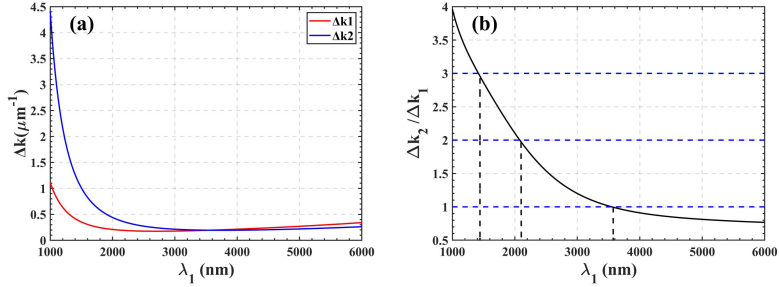

Fig. 3. The ratio of the phase mismatch ($\Delta k_2/\Delta k_1$) between the SFG ($\omega+2\omega=3\omega$) and the SHG ($\omega+\omega=2\omega$) of the PPLN crystal versus the fundamental wavelength ($=2\pi/\omega$) when the temperature is 80 °C. The results show that $\Delta k_2$ is an integer multiple of $\Delta k_1$ when the fundamental wavelength $\lambda_1$=1420nm, 2080nm and 3550nm, respectively.

At the degenerate point the optimal APPLN is actually a PPLN crystal. We will prove this mathematically in the following. When $\Delta k_1 = \Delta k_2$, the objective function (7) is transformed into

$$f(d_1, d_2, \cdots, d_N) = \frac{2}{L^2 \Delta k_1^2} \times \sum_{m=1}^{N} \sum_{n=1}^{m-1} \{\cos(\Delta k_1 m \Delta x) - \cos[\Delta k_1 (m-1) \Delta x]\} \{\cos(\Delta k_1 n \Delta x) - \cos[\Delta k_1 (n-1) \Delta x]\} d_m d_n, \quad (9)$$

Using the identical equation $\left(\sum_{i=1}^{N} a_i\right)^2 = 2\sum_{i=1}^{N}\sum_{j=1}^{i-1} a_i a_j + \sum_{i=1}^{N} a_i^2$, it is possible to transform $f(d_1, d_2, \cdots, d_N)$ into

$$f(d_1, \cdots, d_N) = \frac{1}{L^2 \Delta k_1^2}[g(d_1, \cdots, d_N)^2 - \sum_{m=1}^{N}\{\cos(\Delta k_1 m \Delta x) - \cos[\Delta k_1 (m-1)\Delta x] d_m\}^2], \quad (10)$$

where, $g(d_1, \cdots, d_N) \equiv \sum_{m=1}^{N} \text{Re}(e^{i\Delta k_1 m \Delta x} - e^{i\Delta k_1 (m-1)\Delta x}) d_m$. Since $d_m^2 \equiv 1$, equation (10) can be written as

$$f(d_1, \cdots, d_N) = \frac{1}{L^2 \Delta k_1^2}[g(d_1, \cdots, d_N)^2 - \sum_{m=1}^{N}\{\cos(\Delta k_1 m \Delta x) - \cos[\Delta k_1 (m-1)\Delta x]\}^2] . \quad (11)$$

From above equation, $f(d_1, \cdots, d_N)$ is maximized if and only if $|g(d_1, \cdots, d_N)|$ is maximized. Notice that

$$g(d_1, \cdots, d_N) = \sum_{m=1}^{N} \{\cos(i\Delta k_1 m \Delta x) - \cos[i\Delta k_1 (m-1)\Delta x]\} d_m . \quad (12)$$

Apply triangular equation $\cos\alpha - \cos\beta = -2\sin\frac{\alpha+\beta}{2}\sin\frac{\alpha-\beta}{2}$, then

$$g(d_1,\cdots,d_N) = -2\sum_{m=1}^{N}\sin[\Delta k_1\Delta x(m-\frac{1}{2})]\sin(\Delta k_1\Delta x/2)d_m$$
$$\leq 2\sum_{m=1}^{N}\left|\sin[\Delta k_1\Delta x(m-\frac{1}{2})]\sin(\Delta k_1\Delta x/2)\right| \quad (13)$$

The equal sign of equation (13) holds if and only if the poling direction of the number $m$ crystal domain is $d_m = sign\{-\sin[\Delta k_1\Delta x(m-\frac{1}{2})]\sin(\Delta k_1\Delta x/2)\}$, moreover, $d_m = sign\{-\sin[\Delta k_1\Delta x(m-\frac{1}{2})]\}$ since $\sin(\Delta k_1\Delta x/2) > 0$. Hence the APPLN is degenerated to PPLN with a poling period of $\frac{2\pi}{\Delta k_1}$. It can be understood that since $\Delta k_1 = \Delta k_2$ at the degenerate point, a PPLN with identical poling periods of $\frac{2\pi}{\Delta k_1}$ satisfies phase matching of two processes simultaneously.

For understanding the mechanism of CTHG $d_{eff}$ enhancement at $\lambda_1$ = 1420nm, where $\Delta k_2/\Delta k_1 = 3$, we visualize the local gains of $\lambda_2$ and $\lambda_3$ when they propagate in unpoled and poled nonlinear crystals as shown in Fig.4(a) and (b), respectively. Their accumulated gains are shown Fig.4(c) and (d), respectively. Fig.4(d) indicates that the crystal poled for SHG also present gain for SFG. In details, the accumulated gain of SHG in a single poling period $\Lambda_1$ is continuously increasing, meanwhile that of SFG also have some increasement after one poling period $\Lambda_1$ because of the inversion of nonlinearity after $\frac{\Lambda_1}{2}$ (which is $\frac{3\Lambda_2}{2}$).

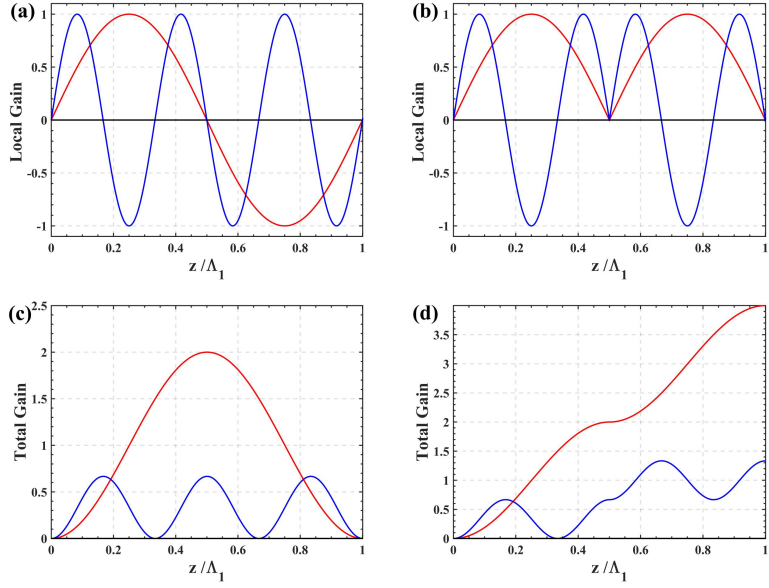

Fig. 4. At the fundamental wavelength of 1420nm, Δk2/Δk1=3, when the crystal is poled with a period of 2π/Δk1, the SHG can be one-order phase matched and the SFG can be third-order phase matched.

The optimal crystal domain distributions, searched by SA algorithm, for $\Delta k_2/\Delta k_1$ = 3, 2, 1 are illustrated in Fig. 5(a), (b), (c), respectively. In Fig. 5(a), the poling length of the first part of the crystal is approximately equal to the period for SHG process, which is the case discussed in above paragraph. This is the reason why the $d_{eff}$ corresponding to the optimal APPLN crystal near the fundamental wavelength of 1420 nm appears a peak in Fig.2.

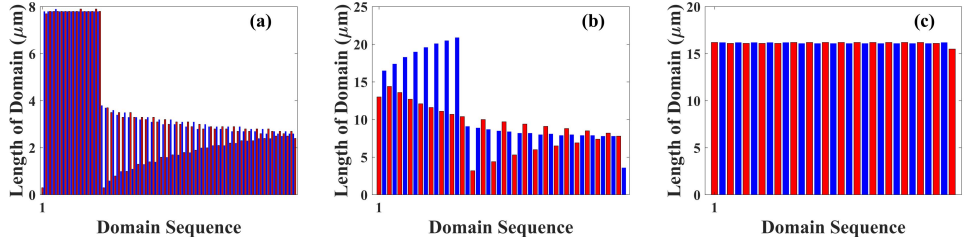

Fig. 5. Optimal CTHG crystal with fundamental wavelength of 1420nm(a), 2080nm(b), and 3550nm(c), with crystal length as 0.5 mm. The blue bar indicates the poling direction downward and the red bar indicates upward, as detailed in Supplementary Material.

## 4. Optimal poling weight analysis

Different from a conventional PPLN for a single second-order nonlinear process with only one reciprocal lattice vector $\Delta k$, an APPLN for CTHG needs two reciprocal lattice vectors, $\Delta k_1$ and $\Delta k_2$. The efficiencies of SHG and SFG are coupled to each other, and each process happens anywhere in the crystal, so there must be an optimal balance in the superlattice design strategy for the ratio between these two processes at each place of the crystal along the propagation direction. Therefore, we define the poling function of the crystal as

$$d_m = sign[a(m)\cos(\Delta k_1 m\Delta x + \phi_1) + b(m)\cos(\Delta k_2 m\Delta x + \phi_2)] , \qquad (14)$$

where $m$ denotes the number of the crystal domain in the beam direction, $a(m)$ and $b(m)$ are the weight factors of the poling period for SHG and SFG, respectively. $\phi_1$ and $\phi_2$ are initial phases. The weighting ratio $c(m) = \dfrac{|a(m)|}{|a(m)| + |b(m)|}$ is defined for the proportion of SHG in the APPLN poling. Thesis [20] gives formula for the optimal distribution of the weight ratio $c(m)$ under certain approximations (see Supplementary Material).

Fig.6 shows the comparison of $d_{eff}$ between the optimal APPLN calculated by Zhang's formula and those by the SA algorithm when the length of the crystal is $L$=0.5mm and 1mm. In Fig.6, the curves by Zhang's formula are very close to our results in most positions, but have obvious deviation in the vicinities of $\lambda_1$=1420nm and $\lambda_1$=3550nm. In these areas the APPLN superlattice arrangements searched by SA have higher $d_{eff}$ than those by Zhang's formula. We will discuss these through analysis and simulation in the following.

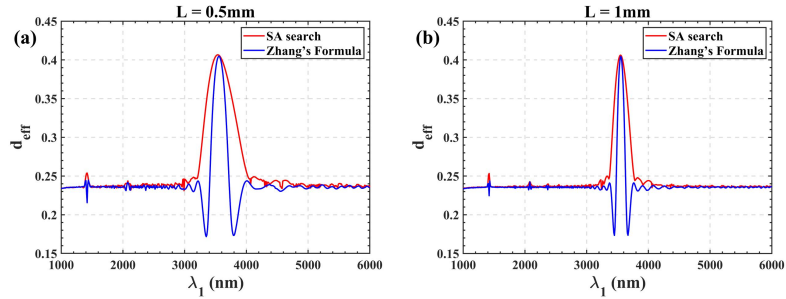

Fig. 6. Comparison of the $d_{eff}$ calculated from Zhang's formula(blue) and those by our SA simulation (red) with crystal length L = 0.5 mm (a) and 1 mm (b).

The basic idea of our calculation is to extract the sum of the terms related to the crystal domain $d_m$ in the objective function (7), denoted as $f_m$, and by calculation we show that $f_m$ can be transformed into the form of the poling function (14). In order to keep the objective function as close to the maximum as possible, $f_m > 0$ is needed to obtain the poling direction of $d_m$. From the expression of $d_m$ at this point, we can derive reversely the distribution curve of the weights $a(m)$ and $b(m)$ corresponding to the actual optimal APPLN.

The objective function (7) can be rewritten as

$$f(d_1, d_2, ..., d_N) = \frac{2}{L^2 \Delta k_1 \Delta k_2} \sum_{m=1}^{N} \sum_{n=1}^{m-1} F_2(m) F_1(n) d_m d_n, \tag{15}$$

where

$$F_1(n) = \cos(\Delta k_1 n \Delta x) - \cos[\Delta k_1 (n-1) \Delta x], \text{ and}$$
$$F_2(m) = \cos(\Delta k_2 m \Delta x) - \cos[\Delta k_2 (m-1) \Delta x].$$

Neglecting the coefficient $\frac{2}{L^2 \Delta k_1 \Delta k_2}$, sum of the terms in (15) associated with $d_m$ is

$$\begin{aligned}
f_m &= d_m d_1 F_2(m) F_1(1) + d_m d_2 F_2(m) F_1(2) + \cdots + d_m d_{m-1} F_2(m) F_1(m-1) \\
&\quad + d_{m+1} d_m F_2(m+1) F_1(m) + d_{m+2} d_m F_2(m+1) F_1(m) + \cdots + d_N d_m F_2(N) F_1(m) \\
&= d_m \left[ F_2(m) \sum_{j=1}^{m-1} F_1(j) d_j + F_1(m) \sum_{i=m+1}^{N} F_2(i) d_i \right].
\end{aligned} \tag{16}$$

Since $\cos\alpha - \cos\beta = -2\sin\frac{\alpha+\beta}{2}\sin\frac{\alpha-\beta}{2}$,

$$\begin{aligned}
F_1(n) &= -2\sin[\Delta k_1 \Delta x (n - \frac{1}{2})] \sin(\frac{\Delta k_1 \Delta x}{2}), \\
F_2(n) &= -2\sin[\Delta k_2 \Delta x (n - \frac{1}{2})] \sin(\frac{\Delta k_2 \Delta x}{2}).
\end{aligned} \tag{17}$$

Then (16) can be written as

$$\begin{aligned}
f_m = d_m \{ &-2\sin[\Delta k_2 \Delta x (m - \frac{1}{2})] \sin(\frac{\Delta k_2 \Delta x}{2}) \times \\
&\sum_{j=1}^{m-1} F_1(j) d_j - 2\sin[\Delta k_1 \Delta x (m - \frac{1}{2})] \sin(\frac{\Delta k_1 \Delta x}{2}) \sum_{i=m+1}^{N} F_2(i) d_i \},
\end{aligned} \tag{18}$$

which can be in the form of a cosine function

$$\begin{aligned}
f_m = d_m \{ &-2\cos(\Delta k_2 \Delta x m - \frac{1}{2}\Delta k_2 \Delta x - \frac{\pi}{2}) \sin(\frac{\Delta k_2 \Delta x}{2}) \times \\
&\sum_{j=1}^{m-1} F_1(j) d_j - 2\cos(\Delta k_1 \Delta x m - \frac{1}{2}\Delta k_1 \Delta x - \frac{\pi}{2}) \sin(\frac{\Delta k_1 \Delta x}{2}) \sum_{i=m+1}^{N} F_2(i) d_i \}.
\end{aligned} \tag{19}$$

If the sign of the $m^{\text{th}}$ domain

$$\begin{aligned}
d_m = \text{sign}\{ &-2\cos(\Delta k_2 \Delta x m - \frac{1}{2}\Delta k_2 \Delta x - \frac{\pi}{2}) \sin(\frac{\Delta k_2 \Delta x}{2}) \times \\
&\sum_{j=1}^{m-1} F_1(j) d_j - 2\cos(\Delta k_1 \Delta x m - \frac{1}{2}\Delta k_1 \Delta x - \frac{\pi}{2}) \sin(\frac{\Delta k_1 \Delta x}{2}) \sum_{i=m+1}^{N} F_2(i) d_i \},
\end{aligned} \tag{20}$$

then $f_m > 0$, and the objective function $f$ is closer to the maximum. Since $m$ is an arbitrary integer between $1 \sim N$, equation (20) presents the solution for every domain in the crystal.

Comparison of (14) and (20) shows that weight factors of the poling period for SHG and SFG are

$$a(m) = -2\sin(\frac{\Delta k_1 \Delta x}{2}) \sum_{i=m+1}^{N} F_2(i) d_i,$$
$$b(m) = -2\sin(\frac{\Delta k_2 \Delta x}{2}) \sum_{j=1}^{m-1} F_1(j) d_j,$$
(21)

and initial phases in (14) are $\phi_1 = \dfrac{-\Delta k_1 \Delta x - \pi}{2}$ and $\phi_2 = \dfrac{-\Delta k_2 \Delta x - \pi}{2}$.

In section 3, the optimal APPLN for the fundamental wavelength from 1000nm to 6000nm were calculated by SA. Applying formula (21), we can derive reversely the distribution of weights $a(m)$, $b(m)$ and $c(m)$ from the optimal APPLN designs at $\lambda_1 = 1000$nm, 3300nm, 3550nm, respectively. The results are shown in Fig.7. The optimized domain distributions of the APPLNs calculated from Zhang's formula and SA algorithm at above three fundamental wavelengths are illustrated in Fig.7.

As shown in Fig.7, the weight ratio curve $c(m)$ at $\lambda_1 = 1000$nm is very smooth and close to that from Zhang's formula (a1), but they significantly deviated from each other (a2), corresponding to different domain distributions as shown in (b2), (c2), respectively. Because the domain distribution in Fig.7(c2) was optimized by SA and the corresponding $d_{eff}$ obtained by (7) is higher than that from Zhang's formula as shown in Fig.6(a) at 3300nm. We think the APPLN designed by us is better than Zhang's formula at this fundamental wavelength. At degenerate point of 3550nm, the solid red line of weight ratio curve $c(m)$ as shown in Fig.7(a3) is a straight line with negative slope, which is also different from the solid blue line, but both are corresponding to almost the same domain distributions in (b3), (c3), respectively. This is because that at degenerate point, $\Delta k_1 = \Delta k_2$, any weight ratio curve gives the result, a PPLN.

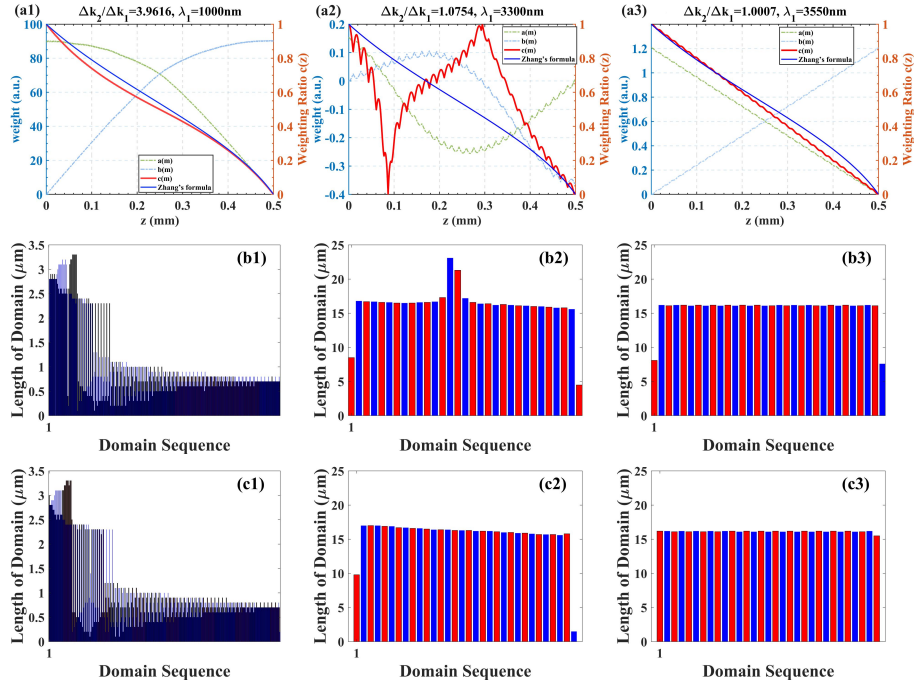

Fig. 7. (a)Weight curves derived reversely from the optimal APPLNs searched by SA at λ1=1000 nm, 3300 nm, 3550 nm, respectively.(b)The APPLN domain distributions corresponding to Zhang's analytical formula with L=0.5mm. (c) The APPLN domain distribution from the SA optimization.

It is easy to expect different $d_{eff}$ curves of different nonlinear materials. We plot the simulation results in Fig.8 for APPLN and aperiodically poled lithium tantalate (APPLT). It shows the curve of APPLT shifts to the short wavelength with degenerate point at 3370nm.

Actually, the method we proposed above is not only suitable for CTHG (SHG+SFG), but also any coupled two $\chi^{(2)}$ processes (with corresponding $\Delta k_1$ and $\Delta k_2$) in a single QPM crystal, such as SHG+DFG, DFG+SFG, etc.

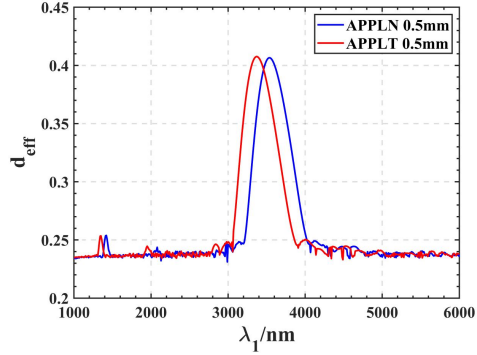

Fig. 8. Values of the effective nonlinear coefficients corresponding to the optimal APPLN crystal (blue) and APPLT crystal (red) for fundamental wavelengths from 1000 to 6000 nm when the crystal length L is 0.5 mm. It can be seen that the effective nonlinear coefficients of the two crystal obey the same distribution law despite the different types of crystal.

## 5. Upper bond of $d_{eff}$

In previous sections, we used a SA algorithm to calculate the "optimal" CTHG crystal for different fundamental wavelengths, however, such a search may be limited to the local optimal solution rather than the global optimal solution. One may naturally ask that how far the local optimal solution searched by the SA method is away from the global optimal.

The direct way to solve this problem is to compute the global optimal solution of the objective function (7). However, this is very difficult. For a crystal with $N$ domains (calculation steps), there are $2^N$ different arrangements, resulting in no guaranteed global optimal searched out in a reasonable time by conventional algorithm. Therefore, we can only use an indirect method to estimate the upper bound of the optimal solution of the objective function (7).

Maximizing objective function can be converted to a binary quadratic programming (BQP):

$$\begin{aligned} &\max x^T J x \\ &s.t. \quad x_i \in \{+1, -1\}, i = 1, 2, ..., N, \end{aligned} \quad (22)$$

where the elements of matrix $J$ are defined by equation (7).

Here, we use the semidefinite programming (SDP) [21] to compute the upper bound of the objective function $f$. The basic idea is to cancel the restriction of independent variable $d_m$ in equation (7) only to be +1 and -1, and "relax" $d_m$ into an N-dimensional vector that satisfies $|x_m|=1$. By applying the semidefinite programming, the original problem can be transformed into

$$\begin{aligned} &\max Tr(JX) \\ &s.t. \quad X_{ii} = 1, i = 1, 2, ..., N, \end{aligned} \quad (23)$$

where $X=[x_1, x_2, ..., x_N]$ is a $N \times N$ semidefinite positive matrix construct by relaxation vector $x_m$.

The problem (23) after relaxation is a polynomial time problem. Its globally optimal can be found in a reasonable time. This is the essential difference between the relaxed problem (23) and the original problem (22). Obviously, the relaxation problem (23) extends the scope of the definition domain of the original problem (22). But if the vectors in problem (23) are restricted to $x_m = (q_m, 0, 0, \cdots, 0)^T$, $q_m \in \{+1, -1\}$, then the relaxation problem (23) degenerates to the original problem (22).

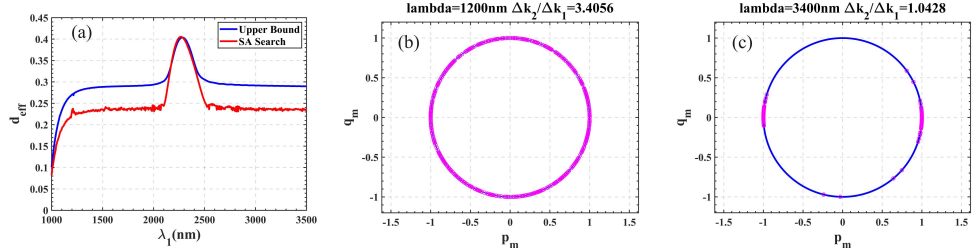

Fig. 9. (a) Maximum $d_{eff}$ obtained by the SA (red) and the theoretical upper bound derived by the SDP (blue). N vectors plotted in two-dimensional plane corresponding to the optimal solution of equation(23) at $\lambda_1$ = 1200 nm (b) and $\lambda_1$ = 3400 nm (c).

For a crystal with $L$=0.5 mm, N=500 when $\Delta x$=1μm. The global optimal solution of the problem (23), which is the theoretical upper bound of the original problem (22), can be derived (see Fig.9). It shows that the optimal solutions derived by the SA are very close to the theoretical upper bound. It is worth noting that the red curve in Fig.9 is different from that in Fig.2 at short wavelengths, since here we use $\Delta x$=1μm, which is not fine enough to get high $d_{eff}$ comparing to $\Delta x$=0.1μm used in Fig.2, but benefit to reduce computing time.

The N-dimensional vectors corresponding to the optimal solutions derived from equation (23) are nearly evenly distributed on a unit circle in a two-dimensional plane when the wavelengths are far away from the degenerate point ($\lambda_1$=3550), i.e. numerical result shows that each $x_m$ corresponding to the solution of problem (23) has only 2 non-zero components, which can be expressed as $x_m = (p_m, q_m, 0, \cdots, 0)^T$, where $p_m^2 + q_m^2 = 1$. Fig.9 (b) illustrates one of the cases, when the fundamental wavelength $\lambda_1$ = 1200 nm. However, there is no such property in the vicinity of the degenerate point (Fig.9 (c)), the reason of which is to be investigated in the future.

## 6. Conclusion

This paper focused on coupled dual second-order nonlinear process and converts the effective nonlinear coefficient into an Ising model, and used the simulated annealing method to find the domain design of the aperiodically poled QPM crystals with the largest effective nonlinear coefficient. Taking the CTHG as an example, the domain distribution of the optimal APPLN is calculated for the fundamental frequency light in the range from 1000nm to 6000nm. The simulation results show that the $d_{eff}$ curve has peaks at $\Delta k_2/\Delta k_1$=1, 2, and 3, respectively, over the entire calculated spectral range. We demonstrate that the optimal APPLN degenerates into a PPLN at the degenerate point ($\Delta k_2 = \Delta k_1$), where the $d_{eff}$ can be as high as 0.41, which is 1.7 times higher than that at the general wavelength. In this paper, we proposed a reconstruction of the weight ratio of the dual nonlinear process at different positions within the crystal based on the optimized APPLN domain distribution. The obtained weight curves were compared with those generated from the analytical approximation formula. The numerical solutions and the approximate analytical curves are quite different on both sides of the degenerate point. We also estimated the theoretical upper bound of $d_{eff}$ using the semidefinite programming, and the results show that the optimal solution obtained by the SA method is close to the theoretical upper bound, especially in the vicinity of the degenerate point. The methods proposed in this paper are applicable to any coupled dual second-order nonlinear process, provide guidance for efficient coupled QPM crystal design, and contributes to the understanding of the dynamics of multiprocess nonlinear frequency conversion.

**Funding.** National Key Research and Development Program of China (2022YFB4601103); Guangdong Basic and Applied Basic Research Foundation (2020B1515120041).

**Disclosures.** The authors declare no conflicts of interest.

**Data availability.** Data underlying the results presented in this paper are not publicly available at this time but may be obtained from the authors upon reasonable request.

See Supplementary Materials for supporting content.

**References**

1. T. Udem, R. Holzwarth, and T. W. Hänsch, "Optical frequency metrology," Nature **416**, 233-237 (2002).
2. S. A. Diddams, K. Vahala, and T. Udem, "Optical frequency combs: Coherently uniting the electromagnetic spectrum," Science **369**, eaay3676 (2020).
3. J. Li, J. Yan, L. Jiang, J. Yu, H. Guo, and L. Qu, "Nanoscale multi-beam lithography of photonic crystals with ultrafast laser," Light: Science & Applications **12**, 164 (2023).
4. B.P. Abbott, *et al*., "GW170817: observation of gravitational waves from a binary neutron star inspiral." Physical review letters **119**, 161101 (2017).
5. T. Bothwell, C. J. Kennedy, A. Aeppli, D. Kedar, J. M. Robinson, E. Oelker, A. Staron, and J. Ye, "Resolving the gravitational redshift across a millimetre-scale atomic sample," Nature **602**, 420-424 (2022).
6. H.-S. Zhong, H. Wang, Y.-H. Deng, M.-C. Chen, L.-C. Peng, Y.-H. Luo, J. Qin, D. Wu, X. Ding, and Y. Hu, "Quantum computational advantage using photons," Science **370**, 1460-1463 (2020).
7. H.-T. Zhu, Y. Huang, H. Liu, P. Zeng, M. Zou, Y. Dai, S. Tang, H. Li, L. You, and Z. Wang, "Experimental mode-pairing measurement-device-independent quantum key distribution without global phase locking," Physical Review Letters **130**, 030801 (2023).
8. R. W. Boyd, *Nonlinear Optics, Third Edition* (Elsevier, 2010).
9. S.-N. Zhu, Y.-Y. Zhu, and N.-B. Ming, "Quasi-phase-matched third-harmonic generation in a quasi-periodic optical superlattice," Science **278**, 843-846 (1997).
10. M. M. Fejer, G. Magel, D. H. Jundt, and R. L. Byer, "Quasi-phase-matched second harmonic generation: tuning and tolerances," IEEE Journal of quantum electronics **28**, 2631-2654 (1992).
11. L. E. Myers, R. Eckardt, M. Fejer, R. Byer, W. Bosenberg, and J. Pierce, "Quasi-phase-matched optical parametric oscillators in bulk periodically poled LiNbO$_3$," JOSA B **12**(11), 2102-2116 (1995).
12. S.-N. Zhu, Y.-Y. Zhu, Y.-Q. Qin, H.-F. Wang, C.-Z. Ge, and N.-B. Ming, "Experimental Realization of Second Harmonic Generation in a Fibonacci Optical Superlattice of LiTaO$_3$," Physical review letters **78**, 2752 (1997).
13. M. Arbore, O. Marco, and M. Fejer, "Pulse compression during second-harmonic generation in aperiodic quasi-phase-matching gratings," Optics Letters **22**(12), 865-867 (1997).
14. C.R. Phillips, and M.M. Fejer, "Efficiency and phase of optical parametric amplification in chirped quasi-phase-matched gratings," Optics letters **35**(18), 3093-3095 (2010).
15. B.-Y. Gu, Y. Zhang, and B.-Z. Dong, "Investigations of harmonic generations in aperiodic optical superlattices," Journal of Applied physics **87**(11), 7629-7637 (2000).
16. U. K. Sapaev and D. T. Reid, "General second-harmonic pulse shaping in grating-engineered quasi-phase-matched nonlinear crystals," Optics Express **13**(9), 3264-3276 (2005).
17. J.-Y. Lai, Y.-J. Liu, H.-Y. Wu, Y.-H. Chen, and S.-D. Yang, "Engineered multiwavelength conversion using nonperiodic optical superlattice optimized by genetic algorithm," Optics express **18**(5), 5328-5337 (2010).
18. Z. Zheng, S. Yang, D. T. Reid, Z. Wei, and J. Sun, "Design of quasi-phase-matching nonlinear crystals based on quantum computing," Frontiers in Physics **10**, 1135 (2022).
19. N. Mohseni, P. L. McMahon, and T. Byrnes, "Ising machines as hardware solvers of combinatorial optimization problems," Nature Reviews Physics **4**, 363-379 (2022).
20. C. Zhang, "Dynamic theoretical study of the coupling phase matching process of nonlinear optical superlattices," PhD thesis of NanJing University, 2004. (written in Chinese)
21. M. X. Goemans and D. P. Williamson, "Improved approximation algorithms for maximum cut and satisfiability problems using semidefinite programming," Journal of the ACM **42**(6), 1115-1145 (1995).
22. Y.-Q. Qin, Y.-Y. Zhu, S.-N. Zhu, and N.-B. Ming, "Quasi-phase-matched harmonic generation through coupled parametric processes in a quasiperiodic optical superlattice," Journal of applied physics **84**(12), 6911-6916 (1998).

23. O. Gayer, Z. Sacks, E. Galun, and A. Arie, "Temperature and wavelength dependent refractive index equations for MgO-doped congruent and stoichiometric LiNbO$_3$," Applied Physics B **91**, 343-348 (2008).
24. A. Lucas, "Ising formulations of many NP problems," Frontiers in physics **2**, 5 (2014).
25. B.-Q. Chen, C. Zhang, C.-Y. Hu, R.-J. Liu, and Z.-Y. Li, "High-efficiency broadband high-harmonic generation from a single quasi-phase-matching nonlinear crystal," Physical Review Letters **115**, 083902 (2015).
26. T. Chen, B. Wu, W. Liu, P. Jiang, J. Kong, and Y. Shen, "Efficient parametric conversion from 1.06 to 3.8 μm by an aperiodically poled cascaded lithium niobate," Optics letters **36**(6), 921-923 (2011).
27. W. Zhu, C. Han, E. Huffman, J. S. Hofmann, and Y. He, "Uncovering Conformal Symmetry in the 3D Ising Transition: State-Operator Correspondence from a Quantum Fuzzy Sphere Regularization," Physical Review X **13**, 021009 (2022).

# NUMERICAL INVESTIGATION OF EFFECTIVE NONLINEAR COEFFICIENT MODEL FOR COUPLED THIRD HARMONIC GENERATION: SUPPLEMENTARY DOCUMENT

## S1. Histogram representation of APPLN crystal poled domains

In this paper, we used histograms to represent the domain distributions of APPLNs, which is intuitive to illustrate the change of the length of the crystal domains along the light propagation direction.

In our simulation, we used evenly spaced steps($\Delta x$) for maximizing the effective nonlinear coefficient($d_{eff}$) by determining the direction of each step. As illustrated in Fig.S1(a), a simple APPLN model contains 14 crystal segments with $\Delta x$=1μm (we used shorter steps in our real simulation). Assuming the result of optimal poling direction is as Fig.S1(a), then the corresponding domain distribution is as Fig.S1(b), and the histogram is Fig.S1(c). The red color indicates that the poled direction is upward and the blue color downward.

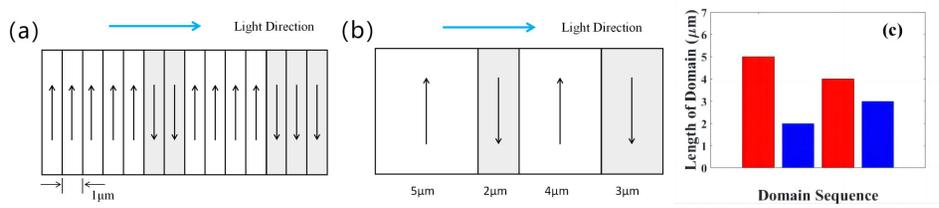

Fig. S1. Histogram representation of the length distribution of APPLN crystal domains. (a) Optimal poled direction of each crystal domain (b) Alternative equivalent representation of the poled direction of crystal domains (c) Histogram representation of the optimal poled direction

## S2. Derivation of coupled third harmony generation small-signal solutions

The coupled third harmony generation (CTHG) process contains both SHG and SFG nonlinear processes, whose coupled wave equations are

$$\frac{dE_1}{dz} = \frac{i\omega d_{33}}{n_1 c} d(z)[E_1^* E_2 \cdot \exp(i\Delta k_1 z) + E_2^* E_3 \cdot \exp(i\Delta k_2 z)] \tag{S1}$$

$$\frac{dE_2}{dz} = \frac{i\omega_2 d_{33}}{2 n_2 c} d(z) E_1^2 \cdot \exp(-i\Delta k_1 z) + \frac{i\omega_2 d_{33}}{n_2 c} d(z) E_1^* E_3 \cdot \exp(i\Delta k_2 z)] \tag{S2}$$

$$\frac{dE_3}{dz} = \frac{i\omega_3 d_{33}}{n_3 c} d(z) \cdot E_1 E_2 \cdot \exp(-i\Delta k_2 z), \tag{S3}$$

where $i$ is imaginary unit, $\Delta k_1$ and $\Delta k_2$ are the phase mismatches of the SHG and SFG processes, respectively. $d_{33}$ is the nonlinear coefficient of the lithium niobate crystal in the 33 direction, $c$ is the speed of light in a vacuum, and $\omega_1$, $\omega_2$, and $\omega_3$ are the frequencies of the fundamental wave, second harmonic wave, and third harmonic wave, respectively. $n_1$, $n_2$, and $n_3$ are the corresponding refractive indices, respectively. $d(z)$ is the poling direction of the domain at position $z$, where $d(z)= +1$ means "up" and $-1$ means "down". In the small-signal approximation, assuming that the nonlinear conversion is not too strong and that the fundamental frequency light $E_1$ remains constant in the crystal, $|E_3|$ is very small compared to $|E_2|$, and the second term on the right side of (S2) can be neglected, $E_2$ can be obtained by

$$E_2(z) = \int_0^z \frac{i\omega_2 d(x)}{2cn_2} E_1^2 e^{-i\Delta k_1 x} dx + E_2(0) = \frac{i\omega_2}{2cn_2} E_1^2 \int_0^z d(x) e^{-i\Delta k_1 x} dx . \tag{S4}$$

Substituting (S4) into (S3) and integrating gives

$$\begin{aligned} E_3(L) &= \int_0^L \frac{i\omega_3 d(z)}{cn_3} E_1 \cdot \left( \int_0^z \frac{i\omega_2 d(x)}{2cn_2} E_1^2 e^{-i\Delta k_1 x} dx \right) e^{-i\Delta k_2 z} dz \\ &= -\frac{\omega_2 \omega_3}{2c^2 n_2 n_3} E_1^3 \int_0^L d(z) e^{-i\Delta k_2 z} dz \int_0^z d(x) e^{-i\Delta k_1 x} dx. \end{aligned} \tag{S5}$$

The effective nonlinear coefficient is expressed as

$$d_{eff} = \left| \frac{2}{L^2} \int_0^L dz\, e^{i\Delta k_2 z} d(z) \times \int_0^z dx\, e^{i\Delta k_1 x} d(x) \right|, \tag{S6}$$

then,

$$E_3(L) = -\frac{\omega_2 \omega_3}{2c^2 n_2 n_3} E_1^3 \frac{L^2}{2} d_{eff} . \tag{S7}$$

Utilizing formula $I = \frac{1}{2} c n \varepsilon_0 |E|^2$,

$$I_3(L) = \frac{I_1^3 \omega_2^2 \omega_3^2 L^4}{4c^6 \varepsilon_0^2 n_1^3 n_2^2 n_3} d_{eff}^2 . \tag{S8}$$

Since $\omega_2 = \frac{4\pi c}{\lambda}$, $\omega_3 = \frac{6\pi c}{\lambda}$, we finally get

$$\eta_{THG} \equiv \frac{I_3(L)}{I_1} = \frac{144\pi^4 |d_{33}|^2 I_1^2 L^4}{c^2 \varepsilon_0^2 \lambda^4 n_1^3 n_2^2 n_3} \left( d_{eff} \right)^2 . \tag{S9}$$

### S3. Zhang's formula

Dr. Chao Zhang [1] derived the poled period distribution of the CTHG optimal APPLN under certain approximations in Section 4.2 of his Ph.D. thesis (C. Zhang, "Dynamic theoretical study of the coupling phase matching process of nonlinear optical superlattices," PhD thesis of NanJing University, 2004. (written in Chinese)), and we restate his results here.

Assuming that the polarization period of the crystal is

$$d(x) = sign\left[ k(x) \cos(\Delta k_1 x) + \cos(\Delta k_2 x) \right] . \tag{S10}$$

A spatially varying weight function $k(x)$ is introduced to regulate the local Fourier coefficients at $x$. Assume that $k(x)$ is a slow-varying function with respect to the crystal length $L$. It can be shown that the local Fourier coefficients at x contribute efficiently to the overall CTHG as

$$\delta A_3 = g_1(x)[1-H_2(x)] + g_2(x)H_1(x). \tag{S11}$$

For optimal crystal domains, there should exist

$$\frac{d\delta A_3}{dg_1} = 0 \tag{S12}$$

in every position $x$. Here,

$$g_1(x) = \begin{cases} E[k(x)], & k(x) \leq 1 \\ k(x)\{E[1/k(x)] + [\frac{1}{k(x)^2} - 1]K[1/k(x)]\}, & k(x) > 1 \end{cases} \tag{S13}$$

$$g_2(x) = \begin{cases} \frac{1}{k(x)}\{E[k(x)] + [k(x)^2 - 1]K[k(x)]\}, & k(x) < 1 \\ E[1/k(x)], & k(x) > 1 \end{cases} \tag{S14}$$

$$H_1(x) = \int_0^x g_1(x)dx \tag{S15}$$

$$H_2(x) = \int_0^x g_2(x)dx \tag{S16}$$

$K(k)$ and $E(k)$ are Legendre complete elliptic integrals of type I and type II, respectively. According to the properties of elliptic integrals

$$\frac{dg_1}{dx} = -\frac{E(k)-K(k)}{k}k' \tag{S17}$$

$$\frac{dg_2}{dx} = -\frac{E(k)-K(k)}{k^2}k'. \tag{S18}$$

Here, $k'=dk/dx$,

$$\frac{dg_1}{dg_2} = -k. \tag{S19}$$

Consider the case $k < 1$,

$$k(x)[1-H_2(x)] - H_1(x) = 0 \tag{S20}$$

$$k'(x)[1-H_2(x)] = kg_2(k) + g_1(k). \tag{S21}$$

Let,

$$F(k) = kg_2(k) + g_1(k) = 2E(k) + (k^2-1)K(k), \tag{S22}$$

it's easy to prove

$$\frac{dF}{dk} = g_2(k). \tag{S23}$$

Combining (S21) and (S22), we get

$$1 - H_2 = \frac{F(k)}{k'}. \tag{S24}$$

Taking the differential of (S24), we get

$$-g_2 = \frac{-Fk'' + F'k'^2}{k'^2}. \tag{S25}$$

It can be simplified as

$$2g_2 = F\frac{k'}{k'^2}. \tag{S26}$$

Integrating both sides results in

$$\frac{2}{F}dF = \frac{1}{k'}dk' \tag{S27}$$

$$\frac{dx}{dk} = \frac{C}{F^2}, \tag{S28}$$

where $C$ is a constant and $C = 1$ from the boundary conditions, so the inverse function of the optimal weight $k(x)$ is

$$x(k) = \int_0^k \frac{1}{F^2(k)}dk, \tag{S29}$$

where

$$F(k) = 2E(k) + (k^2 - 1)K(k). \tag{S30}$$

Define the weight ratio of the optimal APPLN as $c(x) = \frac{k(x)}{1+k(x)}$, then the CTHG optimal weight ratio $c(x)$ can be obtained from the above expression for $x(k)$ as shown

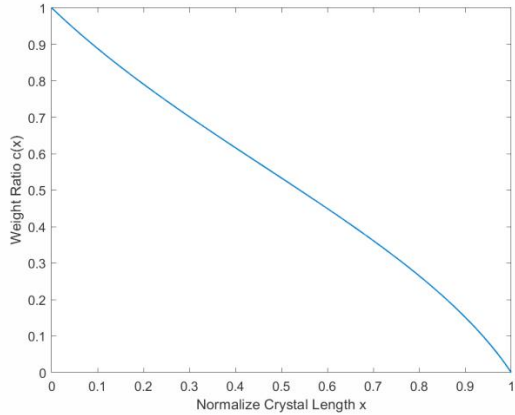

Fig. S2. CTHG optimal weight ratio c(x)

**References**

1. C. Zhang, "Dynamic theoretical study of the coupling phase matching process of nonlinear optical superlattices," PhD thesis of NanJing University, 2004. (written in Chinese)